\pdfoutput=1
\documentclass[aps, reprint, superscriptaddress, nofootinbib, showpacs, floatfix]{revtex4-1}
\usepackage{amsmath, latexsym, amssymb, hyperref, graphicx, color, multirow}
\usepackage[table]{xcolor}
\usepackage{tabularx}
\usepackage[T1]{fontenc}
\usepackage[capitalize]{cleveref}
\definecolor{nicered}{rgb}{.7,.1,.1}
\definecolor{nicegreen}{rgb}{.1,.5,.1}
\definecolor{darkblue}{rgb}{0,0,.5}
\hypersetup{colorlinks, citecolor=nicegreen,linkcolor=nicered, urlcolor=darkblue}

\begin{document}


\title{Minimal renormalizable simplified dark matter model with a pseudoscalar mediator}

\author{Seungwon Baek}
\email{swbaek@kias.re.kr}
\affiliation{School of Physics, Korea Institute for Advanced Study, 85 Hoegiro, Seoul 02455, Korea}

\author{P. Ko}
\email{pko@kias.re.kr}
\affiliation{School of Physics, Korea Institute for Advanced Study, 85 Hoegiro, Seoul 02455, Korea}

\author{Jinmian Li}
\email{jmli@kias.re.kr}
\affiliation{School of Physics, Korea Institute for Advanced Study, 85 Hoegiro, Seoul 02455, Korea}


\begin{abstract}
\noindent
We consider a minimal renormalizable and gauge invariant dark matter (DM) model, in which the singlet fermion 
DM has only axial couplings to a new pseudoscalar mediator. The mixing between the pseudoscalar mediator and the standard model (SM) Higgs boson
induces the interactions between the DM and SM particles. 
The DM candidate in this model can provide the correct thermal relic density and evades all direct detections, 
while it can produce observable signals in indirect detection experiments due to its large annihilation cross section.
A comparative study for DM phenomenology at the LHC is conducted for models with scalar mediators that have 
either scalar or pseudoscalar couplings to SM particles and the DM. 
We find that the three scenarios have distinguishable features in scalar decay branching ratio, DM pair 
production cross section as well as the signal reaches at the LHC. 
The LHC searches for some visible signals related to the scalar sector are also discussed. 

\end{abstract}

\maketitle

\section{Introduction}

The existence of non-baryonic Dark Matter (DM) has been established only by astrophysical observations through 
its gravitational effects~\cite{Ade:2015xua}.  
Since the correct abundance of DM via thermal production could be generically obtained 
if the DM is in the mass range of $\mathcal{O}(100)$ GeV and interacts with SM particles via electroweak force, 
the so-called Weakly-Interacting-Massive-Particle (WIMP)  paradigm has been one of the most interesting 
scenarios for thermal DM.
Given that the DM interactions with the SM particles or among themselves are unknown, effective field theory (EFT) is one viable way to simplify the study of DM phenomenology.
The EFT descriptions~\cite{Goodman:2010yf,Goodman:2010ku,Duch:2014xda} of DM interactions are valid only 
when momentum transfer is much smaller than the mass of the mediator, which is usually not true for DM 
productions at high energy colliders~\cite{Buchmueller:2013dya,Busoni:2013lha,Busoni:2014sya,Busoni:2014haa}, especially since the mediator mass scale is completely unknown.
Simplified DM model frameworks have been used extensively in DM searches at the LHC~\cite{Abdallah:2014hon,Abdallah:2015ter,Abercrombie:2015wmb}. 
Here, the DM is neutral under the Standard Model (SM) gauge group and interacting with the SM particles via the portal of a single particle~\cite{Buckley:2014fba,Harris:2014hga,Harris:2015kda,Yang:2016wrl}. 

However, simplified DM models with a single mediator can often violate the SM gauge symmetry~\cite{Baek:2015lna,Kahlhoefer:2015bea,Englert:2016joy}, thus may become invalid for describing UV-complete models \footnote{Importance of SM gauge symmetry within the DM EFT was pointed out in Ref.~\cite{Bell:2015sza}.}. 
There are growing interests in simplified DM model that respect the gauge symmetry~\cite{Bauer:2016gys,Ko:2016zxg,Baek:2015lna,Bell:2015sza,Baek:2012se,Baker:2015qna,Buschmann:2016hkc,Alves:2015mua}. In particular, the gauge invariant and renormalizable DM model with scalar mediators are constructed in its minimal form~\cite{Baek:2011aa} and two Higgs doublet model (2HDM) extended form~\cite{Bell:2016ekl}. 
Models with pseudoscalar mediators are more interesting, owing to the fact that stringent constraints from DM direct detection can be evaded intrinsically, while being able to explain some anomalies in DM indirect detection~\cite{Ipek:2014gua,Ayazi:2015jij,Kim:2016csm}.
The collider phenomenology of UV-complete DM models with pseudoscalar portal has been studied in Ref.~\cite{Berlin:2015wwa,Fan:2015sza,Goncalves:2016iyg,No:2015xqa,Ghorbani:2016edw}. 

In this work, a minimal renormalizable model with pseudoscalar mediator is proposed (analogy to the model in Ref.~\cite{Ghorbani:2014qpa} which focuses on the DM indirect detection signal). 
Compared to the models in Refs.~\cite{LopezHonorez:2012kv,Esch:2013rta}, the pseudoscalar mediator of this model only has an axial coupling to DM particles. 
We show that there is large portion of parameter space that is consistent with DM constraints while giving measurable predictions in future experiments. 
At the LHC, this model can be searched through signatures both with and without DM in the final state. 
The most remarkable DM signal is produced by recoiling the DM pair against energetic initial state radiated jet, i.e. mono-jet. We will comparatively study these signatures for models with scalar mediators that have either scalar 
or pseudoscalar couplings to SM particles and the DM. 
The pseudoscalar can also produce beyond SM (BSM) signatures without including DM. We will discuss the constraints on the signals of $A \to VV \to (f\bar{f})(f\bar{f})$, $H_0 \to AA \to (f\bar{f})(f\bar{f})$ and $A\to H_0 H_0$ at current stage of the LHC.

\section{Minimal renormalizable model with pseudoscalar mediator}
\label{sec:amodel}
We propose a minimal renormalizable DM model with a pseudoscalar mediator assuming DM $\chi$ is 
a SM singlet Dirac fermion that couples to a pseudoscalar $a$ which is also a SM singlet scalar with 
a negative parity:
\begin{align}
\mathcal{L} &= \bar{\chi}(i \partial \cdot \gamma - m_\chi - i g_\chi a \gamma^5)\chi + \frac{1}{2} \partial_\mu a \partial^\mu a  - \frac{1}{2} m^2_a a^2 \nonumber \\
& - (\mu_a a + \lambda_{Ha} a^2) \left(H^\dagger H -\frac{v_h^2}{2}\right)-\frac{\mu^{\prime}_a}{3!} a^3 - \frac{\lambda_a}{4!} a^4 \nonumber \\
& - \lambda_H \left(H^\dagger H -\frac{v_h^2}{2}\right)^2 ~. \label{lang}
\end{align}
Note that the parity is broken by the dim-3 $\mu_a$ and $\mu^{\prime}_a$ terms. We remove the tadpole for $a$ and assume $\langle a \rangle=0$. 
This model is unique, since the mediator $a$ has a pseudoscalar coupling to the DM $\chi$, 
and scalar couplings to the SM fields through its mixing with the SM Higgs boson (see Eq. (7) below), 
unlike most other renormalizable pseudoscalar mediator models based on 2HDMs 
and its extensions.

The $\mu_a$ term induces the mixing between the pseudoscalar $a$ and the SM Higgs boson $h$ after electroweak symmetry breaking, making two mass eigenstates $H_0$ and $A$:
\begin{align}
H_0 &= h \cos \alpha + a \sin \alpha ~, \\
A &= -h \sin \alpha + a \cos \alpha ~.
\end{align}
So the variables $\lambda_H$, $m^2_a$ and $\mu_a$  in Eq.~\ref{lang} can be expressed by physical parameters in mass eigenstate:
\begin{align}
\lambda_H &= \frac{1}{2 v_h^2} (m^2_{H_0} \cos^2 \alpha + m^2_A \sin^2 \alpha) ~, \\
 m^2_a &= m^2_{H_0} \sin^2 \alpha + m^2_A \cos^2 \alpha ~, \\
 \mu_a &= \frac{\sin\alpha \cos \alpha}{v_h} (m^2_{H_0} - m^2_A)~,
\end{align}
where $v_h$ is the vacuum expectation value of $H$.

Then the interaction Lagrangian of $H_0$ and $A$ with the SM particles and DM will be given by 
\begin{align}
\mathcal{L}_{\text{int}} &=   - i g_\chi (H_0 \sin \alpha + A \cos \alpha) ~ \bar{\chi} \gamma^5 \chi -(H_0 \cos \alpha - A \sin \alpha) \nonumber \\
 & \times  \left[ \sum_f \frac{m_f}{v_h} \bar{f} f - \frac{2m^2_W}{v_h} W^+_\mu W^{-\mu} - \frac{m_Z^2}{v_h} Z_\mu Z^\mu  \right] \label{langint}
\end{align}
The mass eigenstates of scalar fields have only scalar couplings to SM particles and have only axial couplings to DM, so we can expect that such model setup will not lead to any CP-violation effects in the SM. 

On the other hand, the extended Higgs sector could affect the electroweak precision test (EWPT)~\cite{Barger:2007im,Baak:2011ze} by giving extra contributions to the SM gauge boson self-energy. 
Since the new pseudoscalar boson couples to the SM particles only through mixing with the SM Higgs doublet, constraints from the  oblique parameters and the perturbative unitarity bound are exactly the same with the scalar Higgs portal case considered in Ref.~\cite{Baek:2011aa}. Taking $m_{H_0} = 125$ GeV, the measurements exclude the models with scalar mixing angle $\alpha \gtrsim 0.4$. Similar constraint is also obtained from the precision measurements of SM Higgs boson signal strengths at the LHC run-I~\cite{Khachatryan:2014jba,Aad:2015gba}, which indicate $\sin \alpha \lesssim 0.4$~\cite{Robens:2015gla,Cheung:2015dta,Dupuis:2016fda}.  
Moreover, if $m_\chi < m_{H_0}/2$, the stringent limit from the Higgs invisible decay search Br$(H_0 \to \chi \chi)<0.24$~\cite{CMS-PAS-HIG-16-016} requires $g_\chi \sin \alpha \lesssim 0.02 $. 

\section{Dark matter phenomenology}
The measurements of anisotropy of the cosmic microwave background (CMB) and of the spatial distribution of 
galaxies find the relic density for cold non-baryonic matter to be $\Omega h^2 = 0.1198 \pm 0.0026$
~\cite{Ade:2015xua}. 
In order not to overclose the universe, the DM candidate in our model should annihilate effectively into SM particles. There are mainly three different DM annihilation mechanisms in our model framework: (1) DM mass is around the half of $H_0/A$ mass so the annihilation cross section is resonantly enhanced; (2) DM annihilate to SM gauge bosons/heavy fermions especially when $g_\chi \sin 2\alpha$ is large; (3) DM  mass is larger than $H_0$ and/or $A$ so the annihilation cross section can be enhanced by setting large scalar self-coupling. 

The micrOMEGAs~\cite{Belanger:2014vza} is used to calculate the observables in DM phenomenology, 
with the model files for Eq.~(\ref{lang}) generated by Feynrules~\cite{Alloul:2013bka}. 
Taking $H_0$ as the Higgs state with mass of 125 GeV, the model has seven free parameters: 
\begin{align}
m_{A}, ~g_\chi,~\alpha, ~ m_\chi, ~\lambda_{Ha}, ~\mu^{\prime}_a, ~\lambda_a~. 
\end{align} 
In DM annihilation, varying the $g_\chi$ and $\alpha$ can only lead to a total rescaling of the cross section, while its dependence on the $m_\chi$ is more complicate, due to the opening of new annihilation channels with increasing $m_\chi$. Further more, as discussed in Sec.~\ref{sec:amodel}, $\alpha$ should be smaller than 0.4 according to the Higgs precision measurement, but not too small to guarantee sufficient signal rate at collider. So we will choose $g_\chi=1$ and $\alpha=0.3$ for the discussions of this section and scan $m_\chi \in [5,500]~\text{GeV}$. 
The $m_A$ determines position of the pole that is due to resonant enhancement in DM annihilation. Scanning $m_A$ will lead to overlapped peaks in annihilation cross section thus smear out the peak structure. For clarification, we also fix $m_A=400$ GeV. 
The rest of parameters are scanned in the ranges listed as following. 
\begin{align}
  \lambda_{Ha} \in \pm [10^{-3},\sqrt{4 \pi}],  ~\mu^{\prime}_a \in [5,300]~\text{GeV},&~ \lambda_a \in [10^{-3},\sqrt{4\pi}] \label{prange}
\end{align}
We will adopt the exponential scan over the $\lambda_{Ha}$ and $\lambda_a$ in order to have more points with small $\lambda_{i}$, $i=Ha, a$.  That is we define |$\lambda_{i}| = \sqrt{4 \pi}^{R}$ and perform uniform scan over $R$ between [-5.5, 1].

\begin{figure}[thb] \centering
\includegraphics[width=0.47\textwidth]{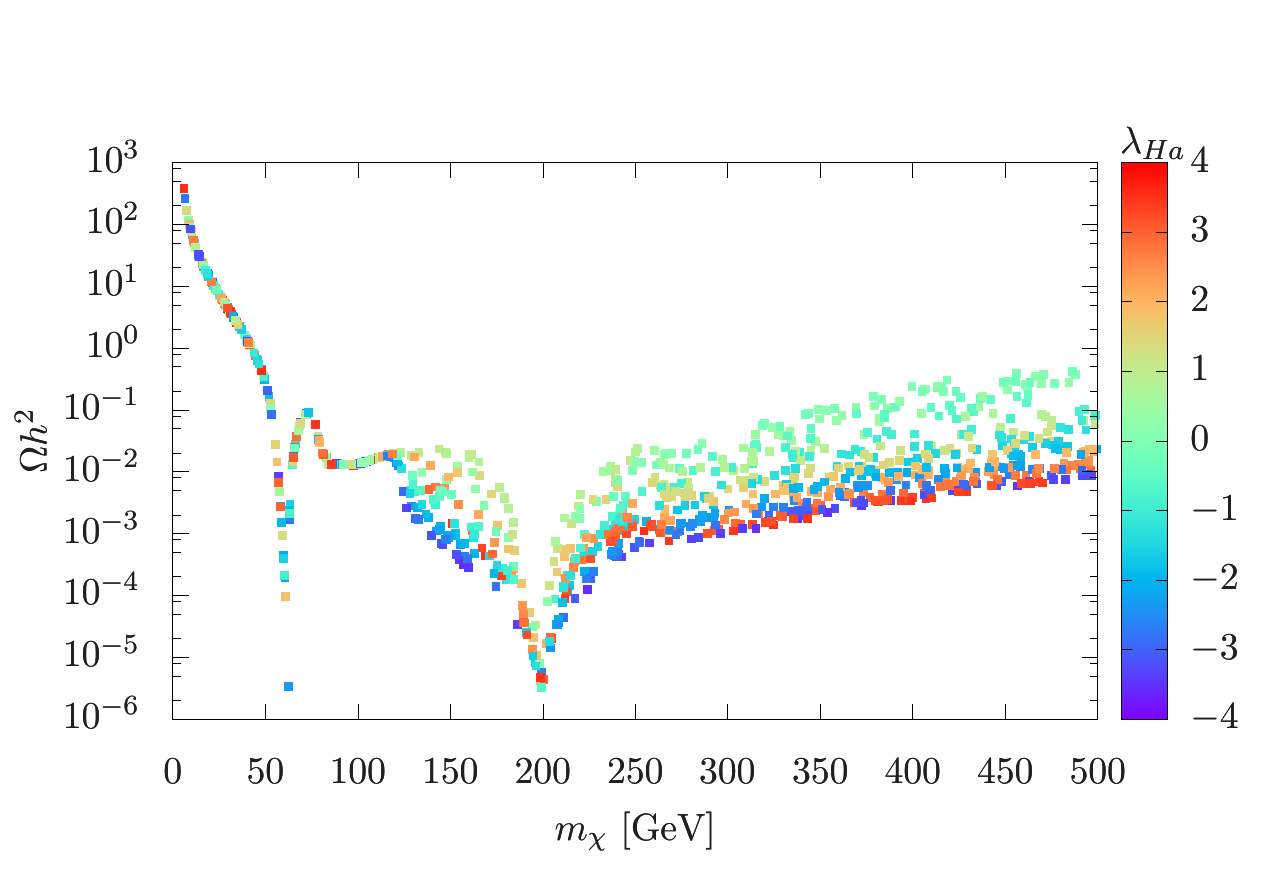}
 \caption{\label{fig:omega} Relic density with varying DM mass, for $m_A =400$ GeV, $g_\chi=1$ and $\alpha=0.3$. Color code indicates the value of $\lambda_{Ha}$.}
\end{figure}

The relic density for models in the chosen parameter space are illustrated in Fig.~\ref{fig:omega}. In the region where DM annihilating into Higgs bosons are kinematically forbidden, $m_\chi$ is the only parameter that control the relic density. The relic density becomes smaller when DM mass is approaching half of the $H_0$ mass. There is also a significant drop at $m_\chi \sim 80$ GeV where the DM annihilating into gauge bosons are opening.  
When $m_\chi \gtrsim m_{H_0/A}$, DM can annihilate into scalar bosons through $H_0/A$ mediation. So the scalar self-couplings are important. Especially, for our parameter choice, $\chi \chi \to A A$ is kinematically disfavored, the relic density is monotonically decreased with increasing |$\lambda_{Ha}$|. 

The DM has been searched actively by many underground experiments through its recoiling against nuclei ~\cite{Akerib:2016lao,Tan:2016zwf}. Following the notations of Ref.~\cite{Kumar:2013iva}, the DM-SM particles interaction can be written in terms of DM bilinear $\mathcal{M}_\chi$, SM bilinear $\mathcal{M}_f$ as well as form factor $F(s,t,u)$ which includes the details of the model and nuclear form factor:
\begin{align}
\mathcal{M} = \mathcal{M}_\chi  \mathcal{M}_f \cdot F(s,t,u). 
\end{align}
In our model, in the limit of low momentum transfer, the DM-SM fermion scattering matrix element is 
\begin{align}
\mathcal{M} & \propto  \mathcal{M}_\chi  \cdot \mathcal{M}_f = -2q^i (\xi_\chi^{\dagger} \hat{S}^i \xi_\chi )\times  \nonumber \\
 & [2m_f (\xi_f^{\dagger} \xi_f) + i\frac{\mu}{m_f} \epsilon^{ijk} q^i v^{j} (\xi_f^{\dagger} \hat{S}^k \xi_f) ]~, \label{sigDD}
\end{align}
where $q^i$ is the momentum transfer, $\xi_{f/\chi}$ are two component spinors for nucleon and DM, $v$ is 
the relative velocity of the dark matter and the target nucleon, $\mu=m_\chi m_f/(m_\chi +m_f)$ is the reduced 
mass of the dark matter-nucleon system.  
Note the $g_\chi$ and $m_A$ dependences are absorbed in $F(s,t,u)$. 
Eq.~(\ref{sigDD}) is showing that the spin-independent (SI) DM-nucleon cross section is suppressed by the $|\vec{q}|^2$ while the spin-dependent cross section is even smaller ($\propto |\vec{q}|^4$). 
The results from the above semi-quantitative estimate can be seen more clearly in the full formula for the SI direct detection cross section,
\begin{align}
\sigma^{SI}_{\chi N} = \frac{2}{\pi} \frac{\mu^4}{m_\chi^2} \lambda_N^2 v^2,
\end{align}
where
\begin{align}
\lambda_N = \frac{g_\chi \sin\alpha \cos\alpha m_N}{v_h} \left(\frac{1}{m_{H_0}^2} -\frac{1}{m_A^2}\right) f_N,
\end{align}
with $N$ denoting nucleon and $f_N \approx 0.47$. 
Assuming the relative velocity between the DM and nuclear is given by the orbital speed of the Sun $\sim \mathcal{O}(10^{-3})$, 
the typical $\sigma^{SI}_{\chi N}$ of our model is around $\mathcal{O}(10^{-6})$ of that in the scalar mediator model~\cite{Baek:2011aa} as also have been justified by comparing the scattering rates of $\hat{\mathcal{O}}_1$ and $\hat{\mathcal{O}}_{11}$ opeartors in Ref.~\cite{Catena:2015vpa}. This means the DM of our model will not leave any signals in direct detection experiments.

However, the s-wave annihilation is still permitted:
\begin{align}
 \mathcal{M}_\chi =\bar{\chi_1} \gamma^5 \chi_2 = -\frac{(E_1 +m_1)(E_2 +m_2)+ \vec{k}^2}{\sqrt{(E_1 +m_1)(E_2 +m_2)}} \xi^{\dagger}_{\chi_1} \xi_{\chi_2}~,
\end{align}
with $\vec{k}$ is the DM momentum. 
So the non-relativistic DM particles that concentrated at the center of galaxies may still have relatively large annihilation cross section. Thus they can be observed in final state of photons~\cite{Atwood:2009ez}, positron/anti-proton~\cite{Picozza:2006nm,Kounine:2012ega} or neutrinos~\cite{Achterberg:2006md}. 

\begin{figure}[htb] \centering
\includegraphics[width=0.47\textwidth]{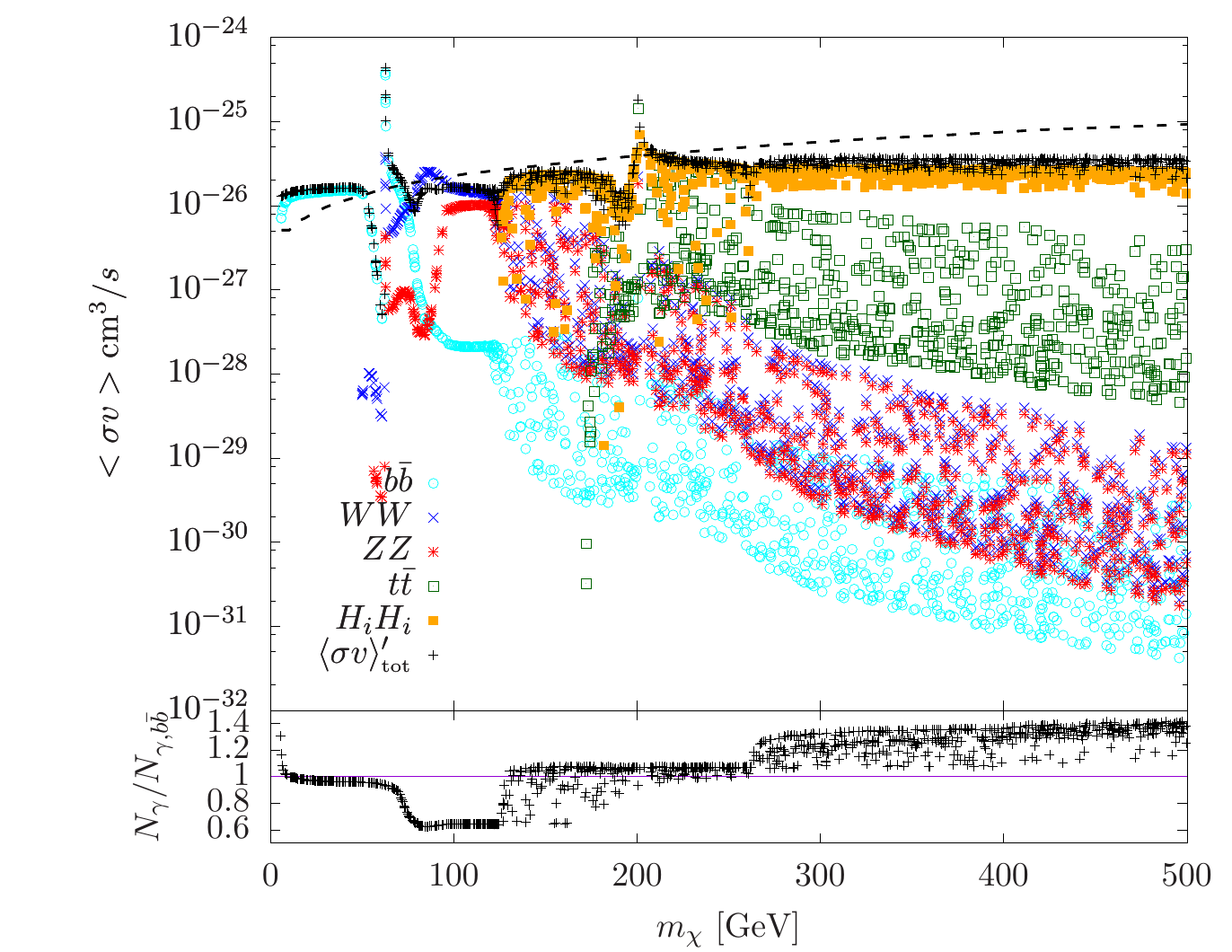}
 \caption{\label{fig:sigmav} Upper panel: The cross sections for different DM annihilation (at rest) channels. The dashed black curve corresponds to the 95\% CL exclusion limit on $b\bar{b}$ channel obtained from Milky Way Dwarf Spheroidal Galaxies with Six Years of Fermi-LAT data~\cite{Ackermann:2015zua}. The weighted total annihilation cross sections are presented by black cross points, which can be compared with the Fermi-LAT data directly. Lower panel: the ratio between the number of photons within energy $E_\gamma \in [1,500]$ GeV per annihilation in our model and in simplified model where DM only annihilates to $b\bar{b}$.  }
\end{figure}

In upper panel of Fig.~\ref{fig:sigmav}, we plot the cross sections for all DM annihilation channels with varying $m_\chi$, where we have chosen appropriate $g_\chi$ such that the correct relic abundance ($\Omega h^2= 0.12$) is obtained for each point in the scanning. 
The exclusion bounds from the Fermi-LAT data are available only for some pure final states, e.g $b\bar{b}$, $\tau^+\tau^-$, $u\bar{u}$ and $W^+ W^-$. 
In order to obtain the Fermi-LAT bound to our model, especially when DM is heavy ($m_\chi \gtrsim 80$ GeV) so that it dominantly annihilates to heavy particles ($W/Z/h/t$), we assume that for a given DM mass the gamma spectra of the $b$ quark and heavy particle final state have similar shape while their normalizations can be different~\cite{Ackermann:2013yva,Agrawal:2014oha}. So we can calculate the weighted total annihilation cross section by 
\begin{align}
\langle \sigma v \rangle ^\prime_{\text{tot}} = \langle \sigma v \rangle_{\text{tot}} \frac{N_{\gamma}}{N_{\gamma, b\bar{b}}}
\end{align}
 where the $\langle \sigma v \rangle_{\text{tot}}$ is the DM total annihilation cross section,
$N_{\gamma}$ is the number of photons within energy $E_\gamma \in [1,500]$ GeV per annihilation for a point in our model and $N_{\gamma, b\bar{b}}$ is the corresponding number in simplified model where DM has the same mass as the point and only annihilates to $b\bar{b}$.
Similar methodology was also pursued in Ref.~\cite{Bringmann:2012vr}. 
We plot the ratio $\frac{N_{\gamma}}{N_{\gamma, b\bar{b}}}$ in the lower panel of Fig.~\ref{fig:sigmav}, from which we can see that the ratio is close to 1 when $\chi\chi \to b\bar{b}$ annihilation is dominant. However, the gauge (Higgs) boson final state can produce less (more) photons in the range $E_\gamma \in [1,500]$ GeV than the $b$ quark final state. 
This also leads to a double enhancement of the ratio at $m_\chi \sim 250$ GeV, where multiple Higgs final state is kinematically opened. 
Then, the weighted total annihilation cross section can be compared to the Fermi-LAT bound on the $b\bar{b}$ final state directly. We can conclude that the Fermi-LAT data from dwarf galaxies can exclude the light DM mass region ($m_\chi < 80$ GeV) as well as the resonant region ($m_\chi \sim m_A /2$), while all of our points are close to the bound and are expected to be discovered/excluded in the near future.  
It has to be noted that this limit will be weakened if our DM particle only constitutes a fraction of the total amount of DM.

\section{LHC phenomenology}
\subsection{Invisible channel: mono-jet}
In this section, we discuss the DM phenomenology at the LHC in terms of decay of scalar, production of DM and current limits from the LHC searches. To show the merit of our model setup, results are presented alongside with those of conventional theoretical frameworks for DM at collider: 
\begin{align}
 \mathcal{L}^{\text{AA}}_{\text{int}} &=   - i g_\chi (a \sin \alpha + A \cos \alpha) ~ \bar{\chi} \gamma^5 \chi  \nonumber \\
 &  -i(a \cos \alpha - A \sin \alpha)  \sum_f \frac{m_f}{v_h} \bar{f} \gamma^5 f  \label{langaa} 
 \end{align}
 \begin{align}
\mathcal{L}^{\text{SS}}_{\text{int}} &=   - g_\chi (H_1 \sin \alpha + H_2 \cos \alpha) ~ \bar{\chi} \chi -(H_1 \cos \alpha - H_2 \sin \alpha) \nonumber \\
 & \times  \left[ \sum_f \frac{m_f}{v_h} \bar{f} f - \frac{2m^2_W}{v_h} W^+_\mu W^{-\mu} - \frac{m_Z^2}{v_h} Z_\mu Z^\mu  \right] \label{langss} 
\end{align}

In the following, we denote the models of Eq.~(\ref{langss}), Eq.~(\ref{langaa}) and Eq.~(\ref{langint}) 
as SS, AA and SA respectively, since they are distinguished by the scalar/axial couplings between SM particles 
and DM. For simplicity, in the discussion of this section $\alpha =0.3$ and $g_\chi=1$ are chosen. And the DM mass is fixed to $m_{\chi}=80$ 
GeV to avoid SM Higgs invisible decay while we keep relatively large DM production cross section. 
The mass of lighter scalar (pseudoscalar) in SS (AA) scenario is chosen as $m_{H_1/a}=125$ GeV 
for comparison purpose.    Then, assuming the $H_2/A$ only decay into SM particles and DM, the only parameter 
relevant in collider phenomenology is $m_{H_2/A}$. This minimal decay width for $H_2/A$ 
(denoted by $A$ hereafter) can be written as 
\begin{align}
\Gamma_{\min}(A) &= \Gamma(A \to \chi \chi) + \Gamma(A \to VV) + \Gamma(A \to ff)  \nonumber \\
  &= \cos^2 \alpha \cdot g^2_\chi \frac{m_A}{8 \pi} (1- \frac{4 m^2_\chi}{m^2_A})^{i/2} \nonumber \\
     &+ {\sin^2 \alpha \cdot \frac{G_\mu m^3_A}{16\sqrt{2} \pi} \delta_V \sqrt{1-4 \frac{m_V^2}{m^2_A}} (1-4 \frac{m_V^2}{m^2_A} +12 \frac{m_V^4}{m^4_A} )} \nonumber \\
     &+ \sin^2 \alpha \cdot (\frac{m_f}{v})^2 \frac{3 m_A}{8 \pi} (1-\frac{4 m^2_f}{m^2_A})^{j/2}~, \label{adecay}
\end{align}
where $(i,j)=(1,3),~(3,3),~(1,1)$ for SA, SS, AA scenarios respectively, $\Gamma(A \to VV) =0$ for AA scenario and $\delta_V=1(2)$ for $Z(W^{\pm})$.

\begin{figure}[h]
\includegraphics[width=0.4\textwidth]{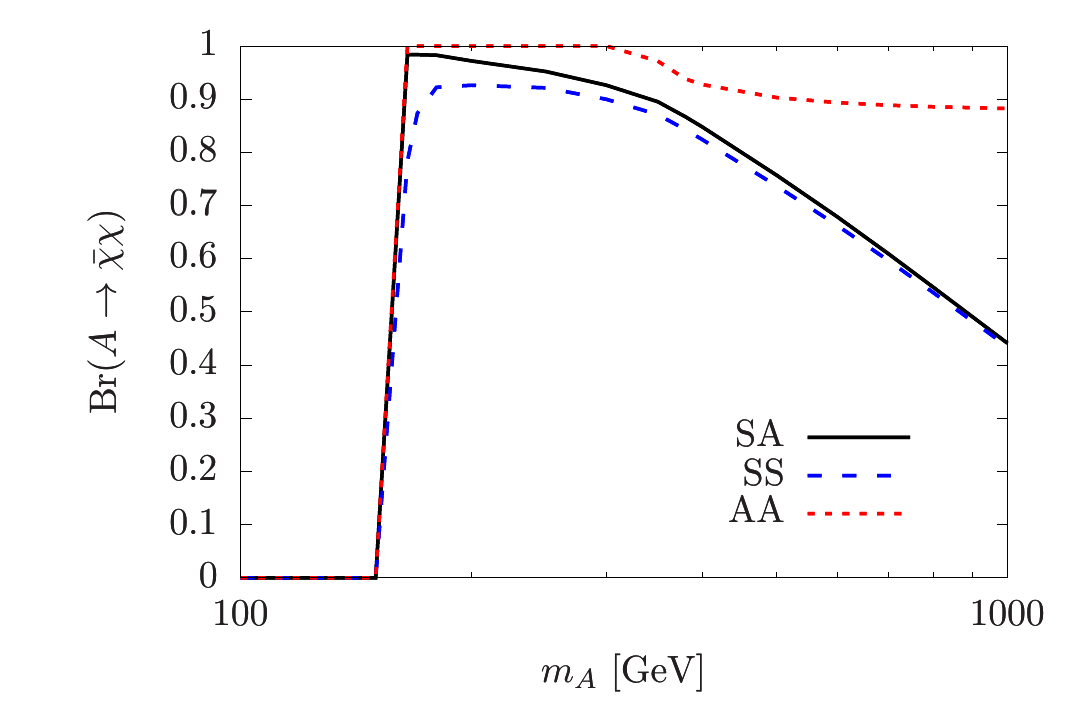}
 \caption{\label{fig:decay} The decay branching ratios of the second scalar boson into DM pair in three models. We have chosen $m_\chi=80$ GeV, $g_\chi=1$ and $\alpha=0.3$.}
\end{figure}

The branching ratios of $A\to \bar{\chi} \chi$ are given in Fig.~\ref{fig:decay}. When the $m_{A}$ is not much larger than $2 m_\chi$, the factor $(1- \frac{4 m^2_\chi}{m^2_A})^{i/2}$ is important. So the Br$(A \to \bar{\chi}\chi)$ of SS scenario is smaller than that of SA scenario. As for $m_A \gg 2 m_\chi$, both scenarios give the similar branching ratios. 
The AA scenario always has the largest Br$(A \to \bar{\chi}\chi)$ because of the absence of $A$-$V$-$V$ coupling. 

The dominant DM production channel at the LHC is gluon-gluon fusion (ggF) through the top quark loop. 
The effective couplings for gluon-gluon-scalar/pseudoscalar after integrating the top quark are
\begin{align}
\mathcal{L}_{\text{scalar}} &= \frac{\alpha_s}{8 \pi} \frac{g_v}{v} \tau [1+(1-\tau) f(\tau)] G^{\mu\nu} G_{\mu\nu} \phi \\
\mathcal{L}_{\text{pseudoscalar}} &= \frac{\alpha_s}{4\pi} \frac{g_v}{v} \tau f(\tau) G^{\mu\nu} \tilde{G}_{\mu\nu} A 
\end{align}
where $\tau = 4 m^2_t/ m^2_{H/A}$, $g_v =\sin \alpha$ and
\begin{equation}
f(\tau) =
  \begin{cases}
    \arcsin^2 \frac{1}{\sqrt{\tau}},       & \quad \tau \geq 1\\
    -\frac{1}{4} (\log \frac{1+\sqrt{1-\tau}}{1-\sqrt{1-\tau}} - i \pi)^2,  & \quad \tau <1 ~.\\
  \end{cases}
\end{equation}

However, the ggF process itself does not produce any observable signals at detectors. Extra energetic jets radiating from either initial state gluon or top quark in the loop can circumvent this issue, which raise the mono-jet signature. The leading order cross section for DM pair production in association with a jet is computed within the FeynRules/MadGraph5\_aMC@NLO~\cite{Alwall:2014hca,Hirschi:2015iia} framework, where the jet is required to have $p_T (j) > 100$ GeV. 
Meanwhile, the higher order corrections to the ggF cross section of Higgs production are found to be quite significant. Using the SusHi program~\cite{Harlander:2012pb}, the NNLO K-factors for Higgs mass $\in [100,500]$ GeV are calculated to be around 2.5. 
So the production cross section for the DM pair associating with a jet is given by the LO cross section in MadGraph5\_aMC@NLO multiplying a universal K-factor of 2.5.

\begin{figure}[h]
\includegraphics[width=0.4\textwidth]{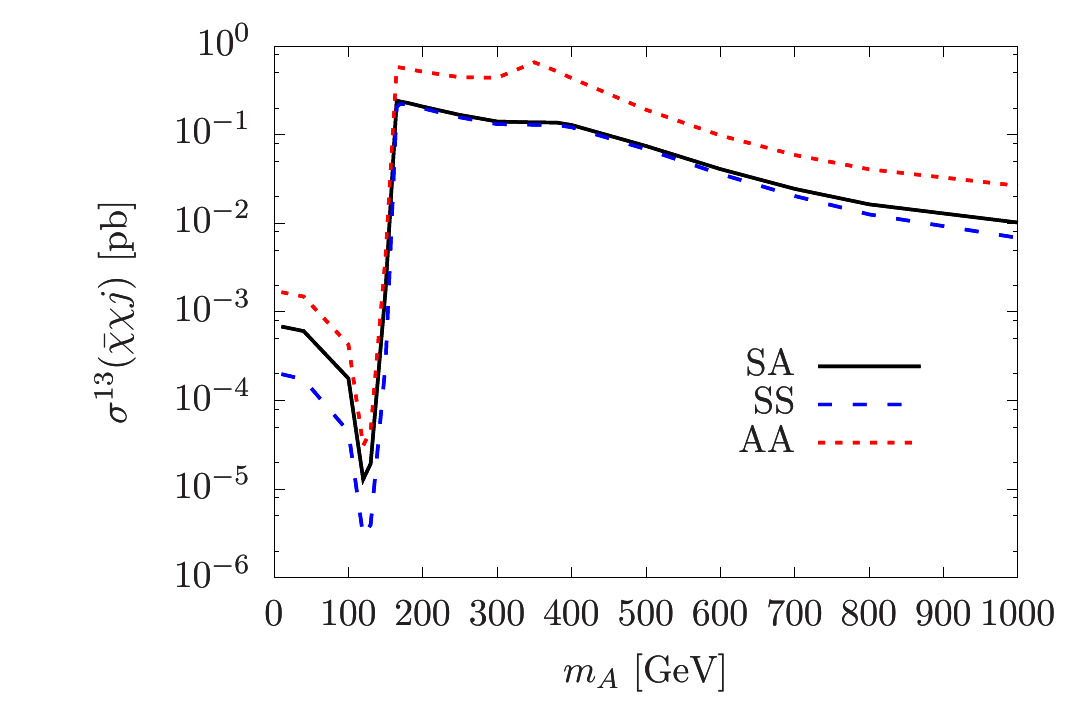}
 \caption{\label{fig:prod} The mono-jet signal production cross section in gluon-gluon fusion channel at the 13 TeV LHC, where the jet is required to have $p_T(j)>100$ GeV. Parameters are chosen as $m_\chi=80$ GeV, $g_\chi=1$ and $\alpha=0.3$. }
\end{figure}

The resulting cross sections for all three scenarios are presented in Fig.~\ref{fig:prod}. 
The contributions of two propagators that mediate the DM production will interference with each other
~\cite{Ko:2016ybp}, leading to different degree of suppressions for different scenarios in the light $m_A$ region. 
In particular, the cross sections drop dramatically when two propagators are close in mass. 
Models with heavier $A$ are more interesting because of their larger production cross section. 
In this region, the DM productions are dominated by the on-shell $A$ production with subsequent decay. 
The interference effect becomes important only for $m_A \gtrsim 700$ GeV, where the on-shell $A$ production 
is kinematically suppressed to some extent. This leads to deviation in the production cross sections of SS and SA 
scenarios.   Note the small bumps around $2m_t$ for all scenarios are from the top quark mass effect. 

The mono-jet signature has been searched by ATLAS collaboration at 13 TeV with integrated luminosity of 3.2 fb$^{-1}$~\cite{Aaboud:2016tnv}. The non detection of the signal could put a constraints on our model parameters. We adopt the CheckMATE2 program~\cite{Dercks:2016npn} to calculate the LHC search constraints on our model, in which the ATLAS mono-jet search has been implemented and validated. 
CheckMATE2 provide the $R^{\max}$-value at the final stage of its analysis, defined as 
\begin{align}
R^{\max} = \max_i \frac{N_i^{\text{model}}}{N_i^{\text{up}}}
\end{align}
where $N_i^{\text{model}}$ and $N_i^{\text{up}}$ is the number of signal events of our model and number of new physics upper limit at 95\% CL in the signal region $i$, respectively. 

\begin{figure}[thb] \centering
\includegraphics[width=0.4\textwidth]{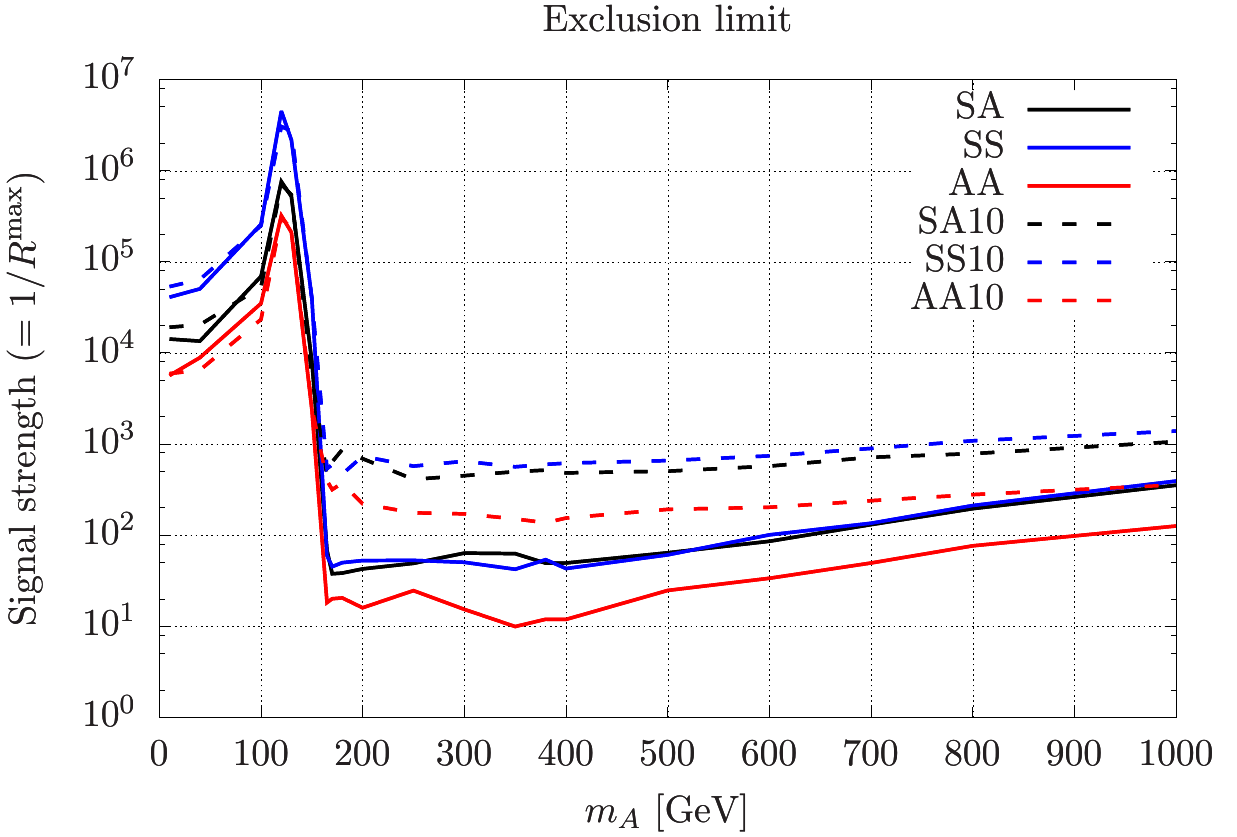}
 \caption{\label{fig:exc} The 95\% CL exclusion limits from the ATLAS mono-jet search at 13 TeV with integrated luminosity of 3.2 fb$^{-1}$ . The dashed curves correspond to models with ten times larger total width of $A$ than $\Gamma_{\min}$ due to the opening of new decay channels.}
\end{figure}

In Fig.~\ref{fig:exc}, we present the LHC search limit with signal strength ($=1/R^{\max}$) which gives the size 
of the cross section that is probable at current stage of the LHC. 
Even in the region of $m_A \gtrsim 2 m_\chi$ where the production cross section is largest, the signal rate is 
at least one order of magnitude below the current reach. 
Note that in this region, since $A$ is mostly on-shell and Br$(A \to \bar{\chi}\chi)$ is already close to 
one, taking larger $g_\chi$ will not enhance the signal rate. 
We would expect higher luminosity of LHC to probe/exclude this region. 
Among three scenarios, the AA scenario has the best search sensitivity. We find that the differences are mainly 
originated from the production rate of mono-jet signals as shown in Fig.~\ref{fig:prod}, while the kinematic 
distributions of final states are similar for all scenarios, i.e. similar cut efficiencies.

In a realistic model, some new decay channels of $A$ might be important, such as $A \to H_0 H_0$. 
This will lead to suppressed production rate of DM pair, meanwhile, the interference effect can become remarkable 
because of the wide width of $A$. In Fig.~\ref{fig:exc}, we also plot the signal reaches for models with ten times 
larger total width of $A$ than $\Gamma_{\min}$ due to the opening of new decay channels. 
In the region with negligible interference, the signal reaches should be one order of magnitude weaker than that 
of models with $\Gamma_{\min}$, e.g. $m_A \in [2 m_\chi, 500~\text{GeV}]$. 
The interference effect is significant when off-shell $A$ contribution is large, e.g. in the regions $m_A > 500$ GeV. 
It shrinks the difference in signal reaches for models with narrow and broad width of $A$, mainly because of the enhancement in production cross section. 
Moreover, the large interference effect can lead to distinguished signal reaches for SS and SA scenarios. 

\subsection{Visible channels}
Our model also predicts BSM signals without DM in the final states. 
In this section, we will focus on the non-DM signals of the SA scenario as we can expect that the exclusion bounds obtained for SA scenario can be directly applied to SS scenario, since their differences only exist in DM sector. But the corresponding bounds in AA scenario could be quite different, due to different production cross section of $A$ as well as the absence of tree level $AZZ/AWW$ couplings. 

According to the Eq.~(\ref{adecay}), 
the heavy pseudoscalar dominantly decays into top quarks and vector bosons apart from the DM pair. 
The process of top quark pair production through the pseudoscalar resonance decay interferes strongly 
with the QCD $t\bar{t}$ background, leading to difficulties in its searches at hadron colliders~
\cite{Dicus:1994bm,Gori:2016zto,Carena:2016npr}. However, the diboson final state may still be detectable. 
To survey the production cross sections of visible signals in our model, we fix $m_\chi=80$ GeV, $g_\chi=1$ and varying $m_A\in [0,1000]$ GeV, $\alpha \in [0,0.3]$, with the rest of parameters scanned in the range as given in Eq.~(\ref{prange}). We note that varying $m_\chi$ and $g_\chi$ which is important in obtaining correct relic density and evading the DM indirect detections will not affect the results in the following discussions much. 

\begin{figure}[thb] \centering
\includegraphics[width=0.47\textwidth]{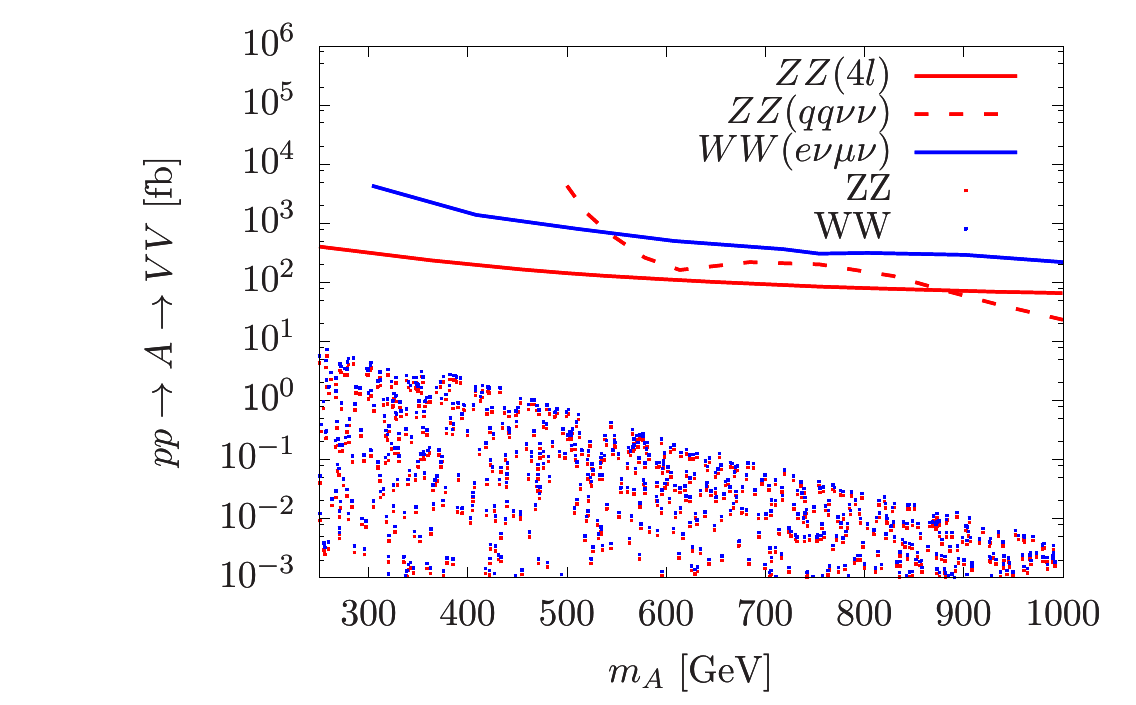}
 \caption{\label{fig:a2vv} Bounds correspond to the LHC searches for two vector boson resonance. The production cross sections of $ZZ~(WW)$ at 13 TeV in our model are shown by red (blue) points. }
\end{figure}

For our parameter choice, the $A\to \chi\chi$ is always important when it is kinematically allowed. So the vector boson pair production cross section is suppressed by $\sim \sin^4 \alpha$, from both $A$ production and decay.  
We calculate the NNLO gluon-gluon fusion $A$ production cross section at 13 TeV by using SusHi and obtain decay branching ratios of $A\to VV$ from micrOMRGAs. 
The results are shown in Fig.~\ref{fig:a2vv}.
At 13 TeV, the vector boson pair production cross section in our model is only around $[0.01,10]$ fb for $m_A \in [200,1000]$ GeV. The ATLAS collaboration searches the high mass diboson resonance in $ZZ \to 4 \ell$~\cite{ATLAS-CONF-2016-079}, $ZZ \to \nu\nu qq$~\cite{ATLAS-CONF-2016-082} and $WW \to e\nu \mu \nu$~\cite{ATLAS-CONF-2016-074}  final states respectively with LHC run-II data. Their exclusion bounds at 95\% confident level (CL) are shown in the Fig.~\ref{fig:a2vv} as well. It can be seen that the signal of vector boson pair production is at least two order of magnitude below the current LHC search sensitivities.

On the other hand, the production rates of scalar pairs ($AA/H_0 H_0$) do not suffer from the $\sin\alpha$ 
suppression as much as those of vector boson pair, because the coupling in scalar to scalar decay is controlled 
by the scalar-scalar mixing and scalar self-couplings:
\begin{align}
\lambda_{AH_0H_0} &= -\mu_a \cos^3 \alpha + 2(3\lambda_H - 2 \lambda_{Ha}) v_h \cos^2\alpha \sin \alpha  \nonumber\\
 & +2\lambda_{Ha} v_h \sin^3 \alpha + (2\mu_a -\mu^{\prime}_a) \cos \alpha \sin^2 \alpha \\ 
 \lambda_{H_0AA} &= -\mu_a \sin^3 \alpha - 2(3\lambda_H - 2 \lambda_{Ha}) v_h \sin^2\alpha \cos \alpha  \nonumber\\
 & -2\lambda_{Ha} v_h \cos^3 \alpha + (2\mu_a -\mu^{\prime}_a) \sin \alpha \cos^2 \alpha \label{eq:lhaa}
\end{align}
They can be either large or small. In the parameter space of our interest, the $H_0 \to AA$ and $A\to H_0 H_0$ can even become dominant. 

\begin{figure}[htb] \centering
\includegraphics[width=0.47\textwidth]{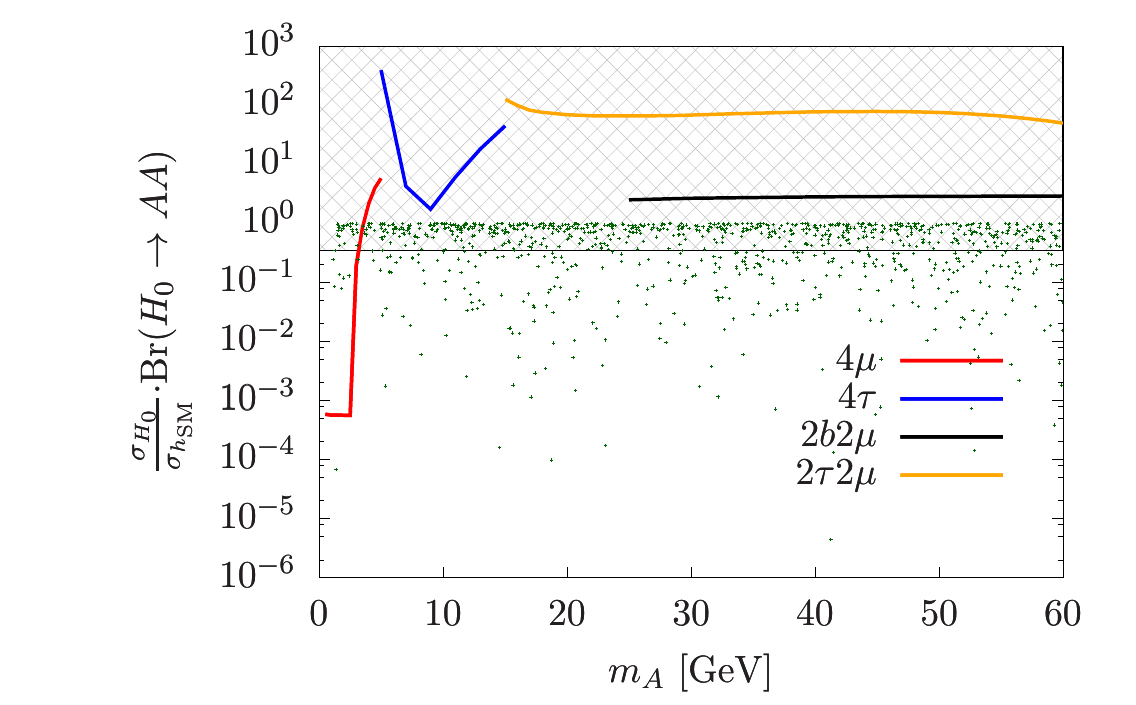}
 \caption{\label{fig:h2aa} Bounds correspond to the LHC searches for light boson pair from the SM Higgs decay. The shaded region is excluded by the Higgs precision measurement. Our models are shown by dark green points. }
\end{figure}

When the $m_A < m_{H_0}/2$, the pseudoscalar pair can be produced from the SM Higgs decay, which will lead to four fermion final states after $A\to f\bar{f}$. The cross section of this process can be quite large. 
Ref.~\cite{Khachatryan:2017mnf} summarizes the recent searches for light bosons from 125 GeV Higgs decay in the final states of $4\mu$, $4\tau$, $2b2\mu$ and $2\tau 2\mu$ at LHC run-I. 
The bounds are presented on the production cross section of each final states normalized to the SM Higgs production cross section. 
In our model, for $m_A \in [0,60]$ GeV, the decay branching fractions of the pseudoscalar are only determined by a single parameter  $m_A$.  So those experimental bounds for different final states can be projected to the same plane, $m_A$ versus $\frac{\sigma_{H_0}}{\sigma_{h_{\text{SM}}}} \cdot \text{Br}(H_0 \to AA)$, where $\frac{\sigma_{H_0}}{\sigma_{h_{\text{SM}}}}= \cos^2 \alpha$. The projected bounds are presented by lines in different colors in Fig.~\ref{fig:h2aa}. 
Further more,  the precision measurements on Higgs coupling strength constrain the BSM Higgs boson decay to be
Br$_{\text{BSM}}\lesssim 34\%$~\cite{Khachatryan:2016vau} as shown by the shaded region of the same figure (it will change slightly for varying $\alpha$).
Finally, we plot the normalized cross section of pseudoscalar pair production of our model by dark-green dots.
We can see from the Fig.~\ref{fig:h2aa} that the $4\mu$ search is quite sensitive to the region $m_A \in [2 m_\mu, 2 m_c]$ where other decay modes are kinematically suppressed while searches for other final states do not have any sensitivities to our model. 
The bound of BSM Higgs boson decay will exclude large portion of the parameter space where the coupling $\lambda_{H_0AA}$ is not suppressed. In the limit of small $\sin\alpha$, Eq.~\ref{eq:lhaa} can be simplified to $\lambda_{H_0 AA} \sim 2 \lambda_{Ha} v_h \cos^3\alpha $. We find the visible points with Br$_{\text{BSM}}\lesssim 34\%$ should have $\lambda_{Ha} \lesssim 0.01$.

\begin{figure}[htb] \centering
\includegraphics[width=0.47\textwidth]{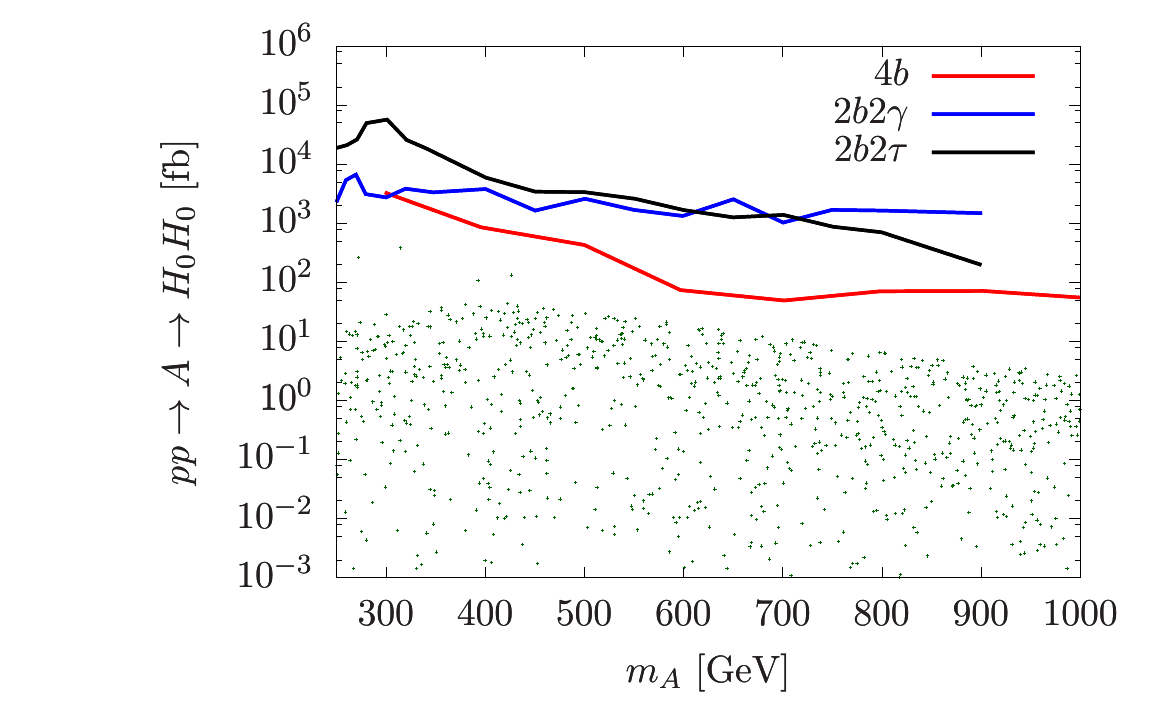}
 \caption{\label{fig:a2hh} Bounds correspond to the LHC di-Higgs searches in different final states. The production cross section of our models at 13 TeV are shown by dark green points. }
\end{figure}

In the region $m_A \in [2m_{H_0},1000~\text{GeV}]$, the $H_0$ pair can be produced through $A$ resonance decay. The cross section of $A$ production is proportional to $\sin^2 \alpha$, while the Br$(A \to H_0 H_0)$ can be large for appropriate choice of parameters in the scalar sector. 
The cross section of resonant $H_0$ pair production from gluon-gluon fusion in our model are shown by dark green points in Fig.~\ref{fig:a2hh}.  The lines in the figure correspond to the 95\% CL LHC searches constraints from $4b$~\cite{ATLAS-CONF-2016-049}, $bb\gamma\gamma$~\cite{CMS-PAS-HIG-16-032} and 
$bb\tau\tau$~\cite{CMS-PAS-HIG-16-029} channels respectively. 
As have been done for Fig.~\ref{fig:h2aa}, the known decay branching ratios of 
$H_0 \to b\bar{b} / \tau \tau / \gamma \gamma$ have been projected out. 
It can been seen that the search for $4b$ final state provides the best sensitivity, and the search for $bb\gamma\gamma$ is better than $4b$ only in the low $m_A$ region. 
For a moderate mass of the pseudoscalar $m_A \sim 600$ GeV, some parameter points are already close to the LHC search limit. We would expect those points can be probed/excluded in the near future when larger data sample is obtained.

\section{Conclusion}

In this paper, we propose a minimal renormalizable and gauge invariant DM model with a pseudoscalar mediator. 
The singlet fermion DM has only axial couplings to the pseudoscalar, while the mixing between the pseudoscalar 
and SM Higgs doublet leads to the interactions of DM and SM fermions and gauge bosons. 
Owing to the s-wave annihilation, the DM relic density can be easily obtained and the DM indirect detection signals are remarkable. 
The momentum suppression in DM-nucleon scattering matrix leads to null signal in all DM direct detection experiments.  

We study the most up-to-date LHC search constraints on signals of the model both with and without DM in the final state. 
The mono-jet signature of our model is studied comparatively with that of models with pure scalar and pure axial couplings between the mediator and SM particles/DMs. 
Three scenarios give different predictions on the decay branching ratio of pseudoscalar/scalar to DM and the DM pair production cross section. As a result, different mono-jet search sensitivities are obtained in different scenarios. 
Among them, the AA scenario has the best search sensitivity at the LHC. And the sensitivity of SA is slightly better than that of AA scenario when the inference effect between two propagators is considerable. 
Due to the $\sin^4 \alpha$ suppression in resonant vector boson pair production, the typical production cross section of resonant vector boson pair is at least two order of magnitude below the current LHC search sensitivity. 
The searches for resonant scalar pairs are more promising. For light $m_{A} \in [0,62.5]$ GeV, the stringent limits on the BSM Higgs boson decay branching ratio obtained from Higgs precision measurements as well as the search for light bosons from 125 GeV Higgs boson decay in $4\mu$ final state exclude very large portion of the parameter space. 
As for heavy $m_{A} \in [250,1000]$ GeV, the production rate is suppressed by $\sin^2\alpha$ while the $A\to H_0 H_0$ can vary freely. A much better sensitivity is obtained for this channel than that for resonant $VV$ channel. Some of the parameter points are less than one order of magnitude away from the current search sensitivity, thus can be probed/excluded in the near future.

{\it Note Added:}
After we submitted this paper on the arXiv.org, we came to learn that the same or similar model 
has been considered in Ref.~\cite{Ghorbani:2014qpa}.  We thank Karim Ghorbani for bringing 
his paper to our attention.

\acknowledgments
JL would like to acknowledge Jong Soo Kim for instructions on CheckMATE.
This work is supported in part by National Research Foundation of Korea (NRF) Research Grant NRF-2015R1A2A1A05001869 (SB,PK, JL), and by the NRF grant funded by the Korea government (MSIP) 
(No. 2009-0083526) through Korea Neutrino Research Center at Seoul National University (PK).

\bibliography{AportalDM}

\end{document}